\documentclass[11pt,a4paper]{article}
\pdfoutput=1
\usepackage{jheppub}
\usepackage{xspace}
\usepackage[tight]{subfigure}
\usepackage{amsmath}
\usepackage{amssymb}
\usepackage{amsfonts}
\usepackage{mathrsfs}
\usepackage{comment}
\usepackage{afterpage}
\usepackage{booktabs}
\usepackage{array}

\usepackage{tabularx}
\newcolumntype{C}[1]{>{\centering\arraybackslash}p{#1}}

\usepackage[title,titletoc]{appendix}

\newcommand{\bea}{\begin{eqnarray}}
\newcommand{\eea}{\end{eqnarray}}
\newcommand{\beanon}{\begin{eqnarray*}}
\newcommand{\eeanon}{\end{eqnarray*}}
\newcommand{\ba}{\begin{array}}
\newcommand{\ea}{\end{array}}
\newcommand{\bd}{\begin{description}}
\newcommand{\ed}{\end{description}}
\newcommand{\bi}{\begin{itemize}}
\newcommand{\ei}{\end{itemize}}
\newcommand{\ben}{\begin{enumerate}}
\newcommand{\een}{\end{enumerate}}
\newcommand{\bc}{\begin{center}}
\newcommand{\ec}{\end{center}}

\newcommand{\eqn}[1]{eq.(\ref{#1})}
\newcommand{\Eqn}[1]{Eq.(\ref{#1})}
\newcommand{\eqns}[2]{eqs.(\ref{#1}--\ref{#2})}
\newcommand{\Eqns}[2]{Eqs.(\ref{#1}--\ref{#2})}

\newcommand{\tbn}[1]{tab.~\ref{#1}}

\newcommand{\fig}[1]{fig.~\ref{#1}}
\newcommand{\Fig}[1]{Fig.~\ref{#1}}

\newcommand{\sect}[1]{sect.~\ref{#1}}
\newcommand{\sects}[2]{sect.~\ref{#1}--\ref{#2}}

\newcommand{\rf}[1]{ref.~\cite{#1}}

\newcommand{\rfs}[1]{refs.~\cite{#1}}  % comma separated list

\newcommand{\MadEvent}{{\tt MadGraph5\_aMC@NLO}\xspace}

\newcommand{\iu}{{i\mkern1mu}}

\title{
Vector boson polarizations in the decay of the Standard Model Higgs
}

\author[a,b]{Ezio Maina}
\affiliation[a]{INFN, Sezione di Torino,\\
Via Giuria 1, 10125 Torino, Italy}
\affiliation[b]{Dipartimento di Fisica, Universit\`a di Torino,\\
Via Giuria 1, 10125 Torino, Italy}
\emailAdd{maina@to.infn.it}
\abstract{
The kinematic distributions of the lepton pairs produced in the decay of the Standard Model Higgs to $ZZ$
and $WW$ are related to the polarization fractions of the virtual vector bosons.
The full amplitude can be decomposed analytically into a sum of polarized terms. Several observables,
in particular the invariant mass of two charged leptons, one from 
each of the bosons, and the lepton angular distribution in the vector boson center of mass are shown to be 
sensitive to the boson polarizations.
}

\begin{document}

\maketitle

\section{Introduction}
\label{sec:Introduction}
Vector Boson (VB) polarizations have attracted a great deal of attention in recent times.
On the one hand, for single boson inclusive production, they can be unambigously predicted in the Standard 
Model. On the other hand, Vector Boson Scattering of 
longitudinally polarized is a crucial probe of the ElectroWeak Symmetry Breaking mechanism.

Since experiments can only observe the boson decay products within a limited subset of the full phase space,
extracting  vector polarizations is not straightforward.

VB polarizations at the LHC have been studied in a number of papers
\cite{Bern:2011ie,Stirling:2012zt,Belyaev:2013nla,Baglio:2018rcu,Baglio:2019nmc,Denner:2020bcz}.

Both CMS  and ATLAS have measured the $W$ polarization fractions 
in the $W+\,$jets \cite{Chatrchyan:2011ig,ATLAS:2012au} channel and in $t\,\bar{t}$ events 
\cite{Aaboud:2016hsq, Khachatryan:2016fky}.
$Z$ polarization fractions at the LHC have been measured in \cite{Aad:2016izn,Khachatryan:2015paa}.
The first
polarization measurement at 13 TeV has been performed by
ATLAS in $WZ$~production \cite{Aaboud:2019gxl}.

In \cite{Ballestrero:2017bxn,Ballestrero:2019qoy,Ballestrero:2020qgv} 
a simple and natural way to define cross sections 
corresponding to vector bosons of definite polarization has been proposed. This allows to use 
polarized templates in fitting the data. Recently \MadEvent has introduced the possibility of generating 
polarized amplitudes \cite{BuarqueFranzosi:2019boy}. 

In this paper, I discuss VB polarizations in the decay of the Standard Model Higgs to $ZZ$ and $WW$, where
only one of the VB can be on mass shell. The process is so simple that the decomposition of the full amplitude
can be performed analitically, yielding a compact and transparent expression.
The polarization fractions in Higgs decay are completely determined as in the case of
single boson inclusive production. Their measurement would provide a test of the SM.
This is a new take  on a process which has been studied 
\cite{Soni:1993jc,Chang:1993jy,Skjold:1993jd,Arens:1994wd,Buszello:2002uu,Choi:2002jk} 
since long before the discovery of the Higgs \cite{Chatrchyan:2012xdj,Aad:2012tfa}.
Precise predictions, including QCD and ElectroWeak NLO corrections, have been been provided in 
\rfs{Bredenstein:2006rh,Boselli:2015aha,Altenkamp:2018bcs}. An analysis including dimension six EFT operators can be
found in \cite{Boselli:2017pef}.

The Higgs decay into four fermions has been studied experimentally, albeit
with limited statistics, in order to determine the spin 
and parity of the Higgs, to set limits on its coupling strength  and anomalous couplings to vector bosons
\cite{Aad:2015rwa,Aad:2015mxa,Sirunyan:2017exp,Sirunyan:2018egh,Sirunyan:2019twz}.

In \sect{sec:WpolDecay} I recall the main ingredients needed in the analysis.
In \sects{sec:ResultsZZ}{sec:ResultsWW}
I discuss a number of observables whose distributions depend on the VB polarizations.
%I show that a new variable, the invariant mass of the two positively charged leptons in the $ZZ$ case and 
%the mass of the two charged
%leptons in the $WW$ case, is useful in separating the different polarizations.
Finally, in \sect{sec:ResultsCuts} I discuss how the distributions are modified in the presence of leptonic cuts
in a simple LHC-like framework.

\section{Vector boson polarizations and angular distribution of its decay products}
\label{sec:WpolDecay}
Let us consider an amplitude in which a weak vector boson decays to a final state fermion pair.
In the Unitary Gauge, it can be expressed as

\begin{equation}\label{eq:Mlep}
\mathcal{M} = \mathcal{M}_{\mu} \frac{i}{k^2 - M^2 + i \Gamma M}\left(-g^{\mu\nu}+\frac{k^{\mu}
k^{\nu}}{M^2}\right) \, J^\mu_h(f,f^\prime)\,,
\end{equation}
where 
\begin{equation}\label{eq:J}
J_\mu^h(f,f^\prime) = \left[ {-i\,g \, \bar{\psi}_{f}^h \,\, \gamma_{\mu}\,
\left(c_L P_L + c_R P_R \right)\, \psi_{{f}^\prime}^h } \right].
\end{equation}
$M$ and $\Gamma$ are the vector boson mass and width, respectively, while
$c_R$ and $c_L$ are the right and 
left handed couplings of the fermions to the $W^+(Z)$, as shown in \tbn{table:couplings},
and $h$ denotes the chirality of the fermion.

The polarization tensor, even when $k^2 \neq M^2$, can be expressed in terms of four polarization vectors 
\cite{Kadeer:2005aq}:
\begin{equation}
-g^{\mu\nu} + \frac{k^{\mu}k^{\nu}}{M^2} = \sum_{\lambda = 1}^4 \varepsilon^{\mu}_\lambda(k)
\varepsilon^{\nu^*}_{\lambda}(k)\,\,.
\label{eq:polexpansion}
\end{equation}

\begingroup
\setlength{\tabcolsep}{10pt} % Default value: 6pt
\renewcommand{\arraystretch}{1.5} % Default value: 1
\begin{table}[th]
%\vspace{0.15in}
\begin{center}
%\hspace*{-2mm}
\begin{tabular}{|C{1.5cm}|C{3.2cm}|C{3.2cm}|C{3.2cm}|}
\hline
     & \bf $c_L$    & \bf $c_R$  & $g_{HVV}$ \\  
\hline
W    &  $1/(s \,\sqrt{2})$  & 0   & $M_W/s$ \\ 
\hline
Z    &  $(I^3_{W,f} - s^2\, Q_f)/(s \, c)$  &  $ - s \, Q_f/c$   & $M_Z/(s\,c)$ \\
\hline
\end{tabular}
\end{center}
\caption{
Weak couplings. $c =  \cos\theta_W = M_W/M_Z$ , $s  = \sin\theta_W $}
\label{table:couplings}
\end{table}
\endgroup

In a frame in which the off shell vector boson propagates along the $(\theta_V,\phi_V)$ axis,
with three momentum $\kappa$,
energy $E$ and invariant mass $\sqrt{Q^2}=\sqrt{E^2-\kappa^2}$, the polarization vectors read:
\begin{align}
\varepsilon^{\mu}_{L} &= \frac{1}{\sqrt 2}
(0,\cos\theta_V\cos\phi_V +  \iu\sin\phi_V , \cos\theta_V\sin\phi_V - \iu\cos\phi_V, - \sin\theta_V)\quad 
\,\,\textrm{(left)} \,,\nonumber \\
\varepsilon^{\mu}_{R} &= \frac{1}{\sqrt 2}
(0,- \cos\theta_V\cos\phi_V +  \iu\sin\phi_V , - \cos\theta_V\sin\phi_V - \iu\cos\phi_V, \sin\theta_V)\,\,
 \textrm{ (right)} \,,\\
\varepsilon^{\mu}_{0} &= 
(\kappa,E \sin\theta_V\cos\phi_V,E \sin\theta_V\sin\phi,E \cos\theta_V)
/\sqrt{Q^2}\,\,\qquad \textrm{(longitudinal)}\,, \nonumber \\
\varepsilon^{\mu}_{A} &= \sqrt{\frac{Q^2 - M^2}{Q^2\,M^2}}
(E, \kappa \sin\theta_V\cos\phi_V,\kappa \sin\theta_V\sin\phi_V,\kappa \cos\theta_V)\quad
 \textrm{(auxiliary)}\,.
\nonumber
\end{align}
On shell, the auxiliary polarization is zero and the longitudinal polarization reduces to the usual expression. 

Consider the decays $H \rightarrow Z^* Z^* \rightarrow \mu^+ \mu^- e^+ e^-$ and  
$H \rightarrow W^{-*} W^{+*} \rightarrow \bar{\nu}_\mu \mu^- e^+ \nu_e$ in the center of mass of the 
Higgs. At lowest order the decay is described by a single, double resonant, diagram.
The corresponding amplitude  is
\begin{align}
\label{eq:MH2ZZ}
\mathcal{M}_{h_1 h_2} &=- i \,g_{HVV} 
J_\rho^{h_1}(f_1,f_1^\prime) \,
 \frac{\sum\varepsilon^{\rho}_{\lambda_1}(k_1)
\varepsilon^{*\mu}_{\lambda_1}(k_1)}{k_1^2 - M^2 + i \Gamma M} \,
g_{\mu\mu^\prime}\,
 \frac{\sum \varepsilon^{\mu^\prime}_{\lambda_2}(k_2)
\varepsilon^{*\nu}_{\lambda_2}(k_2)}{k_2^2 - M^2 + i \Gamma M} \,
J_\nu^{h_2}(f_2,f_2^\prime) \\
=  - i&\, g_{HVV} \frac{ J^{h_1}_\rho(f_1,f_1^\prime) \,
\left( \varepsilon^{\rho}_{0}(k_1)   \varepsilon^{*\nu}_{0}(k_2)\,  \varepsilon_{0}(k_1) \cdot 
\varepsilon_{0}(k_2) 
+\varepsilon^{\rho}_{L}(k_1)   \varepsilon^{*\nu}_{R}(k_2)  + \varepsilon^{\rho}_{R}(k_1)   
\varepsilon^{*\nu}_{L}(k_2)  \right) \,
J_\nu^{h_2}(f_2,f_2^\prime) }
{\left(k_1^2 - M^2 + i \Gamma M\right)\left(k_2^2 - M^2 + i \Gamma M\right)}. \nonumber 
\end{align}

Notice that, since $M_H < 2\, M_V$, the double pole approximation of 
refs.\cite{Aeppli:1993cb,Aeppli:1993rs,Denner:2000bj,Billoni:2013aba,Biedermann:2016guo} 
is not applicable.

Defining the decay amplitudes of the Vector Bosons as
\begin{equation}
\label{eq:Mdec}
\mathcal{M}^\mathcal{D}_{\lambda,h} (i) = J_\mu^{h}(f_i,f_i^\prime) \, \varepsilon^\mu_{\lambda}
\end{equation}
and using $\varepsilon_{R/L}^* = -\varepsilon_{L/R}$ one obtains
\begin{equation}
\label{eq:MH2ZZfinal}
\mathcal{M}_{h_1 h_2} =- ig_{HVV} 
\frac{ f_0 \,\mathcal{M}^\mathcal{D}_{0,h_1}(1)  \mathcal{M}^\mathcal{D}_{0,h_2}(2)\,  
\varepsilon_{0}(k_1) \cdot \varepsilon_{0}(k_2)
 - f_L \,\mathcal{M}^\mathcal{D}_{L,h_1}(1)  \mathcal{M}^\mathcal{D}_{L,h_2}(2) 
- f_R \,\mathcal{M}^\mathcal{D}_{R,h_1}(1)  \mathcal{M}^\mathcal{D}_{R,h_2}(2)}
{\left(k_1^2 - M^2 + i \Gamma M\right)\left(k_2^2 - M^2 + i \Gamma M\right)}
\end{equation}
where we have introduced factors $f_0, f_L, f_R$ to keep track from which vector polarization
each term in the final result originates. $f_0, f_L, f_R$  are equal to one in the Standard Model,
and one can envisage to measure them experimentally as a test of the SM.

The decay amplitudes of the Vector Bosons depend on their polarization and the fermion chirality, which we 
denote as $+/-$.
In the rest frame of the $f {f^\prime} $ pair, they are:
\begin{align}
\label{eq:polamp0}
\mathcal{M}^\mathcal{D}_{0,-} &= ig\,c_L\,2E \,\sin\theta , & \mathcal{M}^\mathcal{D}_{0,+} &= 
ig\,c_R\,2E \,\sin\theta \,, \\
\label{eq:polampL}
\mathcal{M}^\mathcal{D}_{L,-} &= ig\,c_L \sqrt{2} E \left(1 - \cos\theta\right) e^{-i\phi} , & 
\mathcal{M}^\mathcal{D}_{L,+} &= -ig\,c_R \sqrt{2} E \left(1 + \cos\theta\right) e^{-i\phi} \,, \\
\label{eq:polampR}
\mathcal{M}^\mathcal{D}_{R,-} &= ig\,c_L \sqrt{2} E \left(1 + \cos\theta\right) e^{i\phi} , & 
\mathcal{M}^\mathcal{D}_{R,+} &= -ig\,c_R \sqrt{2} E \left(1 - \cos\theta\right) e^{i\phi} \,,
\end{align}
 where $(\theta, \phi)$ are polar and azimuthal angles of the positively charged lepton
 (antineutrino in the $W^-$ case),  relative to the 
 boson direction in the laboratory frame. Notice that, if the boson propagates 
 in the  negative $z$ direction $\phi \rightarrow -\phi$.
 If $Q_i^2$ ($i = 1,2$) are the invariant masses squared of the two fermion pairs, $E$ in 
 \eqns{eq:polamp0}{eq:polampR} is equal $\sqrt{Q_i^2}/2$.
 For massless leptons, the decay amplitude for
 the auxiliary polarization is zero because $\varepsilon^{\mu}_{A}$ is proportional to
 the four--momentum of the virtual boson.
 \Eqns{eq:polamp0}{eq:polampR} show that each polarization is uniquely associated with a specific 
 angular distribution of the charged lepton, even when the $V$ boson is off mass shell and the notion of a 
 vector boson with definite polarization is ill--defined.
 
%
%\begin{figure}[!tbh]
%\centering
%\subfigure[{Integrand $I_0$.}\label{fig:Asq}]
%{\includegraphics[scale=0.5]{Figs/Asq.png}}
%\subfigure[{Integrand $I_1$.}\label{fig:Asq_epseps}]
%{\includegraphics[scale=0.5]{Figs/Asq_epseps.png}}\\
%\subfigure[{Integrand $I_2$.}\label{fig:Asq_epsepssq}]
%{\includegraphics[scale=0.5]{Figs/Asq_epsepssq.png}}
%\caption{Shapes of the $Q_1, Q_2$ dependent parts of $\mathcal{M}^2$.}\label{fig:diag}
%\end{figure}
%

The squared amplitude, summed over fermion polarizations, becomes: 
\begin{equation}
\label{eq:interfpol}
\begin{split}
& \mathcal{M}^2 = 
\frac{4\, g_{HVV}^2 \,Q_1^2 \,Q_2^2}{ \left(\left(Q_1^2 - M^2 \right)^2 +  \Gamma^2 M^2\right)  
 \left(\left(Q_2^2 - M^2 \right)^2 +  \Gamma^2 M^2\right)}
  \biggl[    \biggr. \\
&  \quad +  f_L^2\, \biggl(\,c_R^4 \left(1 + \cos\theta_1\right)^2 \left( 1 + \cos\theta_2\right)^2 
  +   c_L^2 c_R^2  \left( 1 + \cos\theta_1\right)^2 \left( 1 - \cos\theta_2\right)^2     \\
&    \qquad    \qquad + c_L^2 c_R^2 \left(1 - \cos\theta_1\right)^2 \left( 1 + \cos\theta_2\right) ^2 
      + c_L^4 \left( 1 - \cos\theta_1\right)^2 \left( 1 - \cos\theta_2\right) ^2\biggr)          \\
&     + f_R^2\, \biggl(\,c_L^4 \left( 1 + \cos\theta_1\right)^2 \left( 1 + \cos\theta_2\right) ^2 
   + c_L^2 c_R^2  \left( 1 + \cos\theta_1\right) ^2 \left( 1 - \cos\theta_2\right) ^2    \\
&    \qquad    \qquad +   c_L^2 c_R^2 \left( 1 - \cos\theta_1\right)^2 \left( 1 + \cos\theta_2\right) ^2 
      + c_R^4 \left( 1 - \cos\theta_1\right) ^2 \left( 1 - \cos\theta_2\right) ^2 \biggr)         \\
&   + 4\, K^2 f_0^2\, (c_L^2 + c_R^2)^2 \sin^2\theta_1 \sin^2\theta_2  \\ 
&    - 4\,  K\, f_0\, \biggl( \,f_L \,\left( \,c_R^4 \left( 1 + \cos\theta_1\right)\left( 1 + \cos\theta_2\right)   + 
          c_L^4 \left( 1 - \cos\theta_1\right)  \left( 1 - \cos\theta_2\right)   \right.\\
&     \qquad   \qquad  \qquad \left.  -  c_L^2 c_R^2 ( \left( 1 + \cos\theta_1\right) \left( 1 - \cos\theta_2\right)  + \left( 1 - \cos\theta_1
\right) 
         \left( 1 + \cos\theta_2\right) ) \right )             \\ 
&    \qquad \qquad +  f_R \,\left( \,c_L^4 \left( 1 + \cos\theta_1\right) \left( 1 + \cos\theta_2\right)  + 
          c_R^4 \left( 1 - \cos\theta_1\right)  \left( 1 - \cos\theta_2\right)   \right.\\
&    \qquad    \qquad   \qquad  \left. -   c_L^2 c_R^2 (\left( 1 + \cos\theta_1\right) \left( 1 - \cos\theta_2\right)  +\left(1 - \cos\theta_1 
\right) 
\left( 1 + \cos\theta_2\right) ) \right) \\
&   \qquad   \qquad  \qquad  \biggr) \sin\theta_1 \,\sin\theta_2 \,\cos\phi \\
&   + 2\,f_L \,f_R \, (c_L^2 + c_R^2)^2 \,\sin^2\theta_1 \,\sin^2\theta_2  \,\cos(2\,\phi)   \biggl. \biggr].
%&   \qquad   \qquad  \qquad  \biggr) \sin\theta_1 \,\sin\theta_2 \,\cos\phi
%  + 2\,f_L \,f_R \, (c_L^2 + c_R^2)^2 \,\sin^2\theta_1 \,\sin^2\theta_2  \,\cos(2\,\phi)  \, \biggl. \biggr].
\end{split} 
\end{equation}
where $\phi = \phi_1 - \phi_2$.

K denotes the product of longitudinal polarization vectors:
\begin{equation}
K = \varepsilon_{0}(k_1) \cdot \varepsilon_{0}(k_2) = 
\frac{M_H^2 -Q_1^2 -Q_2^2}{\sqrt{4\, Q_1^2\, Q_2^2}},
\end{equation}
since, in the Higgs rest frame we have
\begin{equation}\label{eq:kinematics}
E_1 = \frac{m_H^2 + Q_1^2- Q_2^2 }{2 M_H}, \quad E_2 =m_H - E_1, \quad  \kappa_{1,2} = 
\frac{\sqrt{(m_H^2 + Q_1^2- Q_2^2)^2 -4 m_H^2Q_1^2} }{2 M_H}.
\end{equation}

$K$ is larger when the invariant masses of the virtual vector bosons are small and, therefore, longitudinal 
polarizations yield a larger fraction of soft fermion pairs.

The interference terms in \eqn{eq:interfpol} cancel when the squared
amplitude
is integrated over the full range of the angle $\phi$, or,
equivalently, when the charged lepton can be observed for any value of $\phi$. 

Notice that the azimuthal modulation depends quite strongly on the polar 
angles of the two decays. The amplitude of the oscillation is maximal when both $\theta_1$ and $\theta_2$
are equal to $\pi/2$ and becomes zero when either angle is zero or $\pi$. 

\begin{figure}[!tbh]
\centering
\subfigure{\includegraphics[scale=0.27]{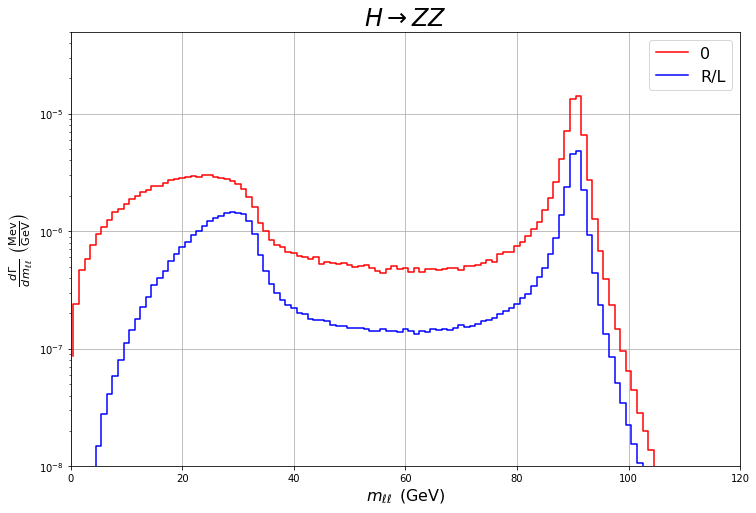}}
\subfigure{\includegraphics[scale=0.27]{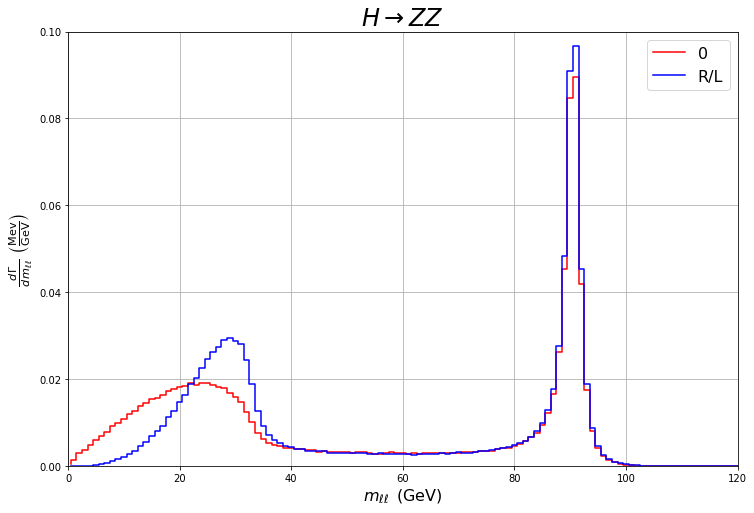}}
\caption{Distribution of the invariant mass of the $\ell^-\ell^+$ pairs. The curves on the right are normalized 
to unit integral.}\label{fig:Zmass}
\end{figure}

\begin{figure}[!tbh]
\centering
\subfigure{\includegraphics[scale=0.3]{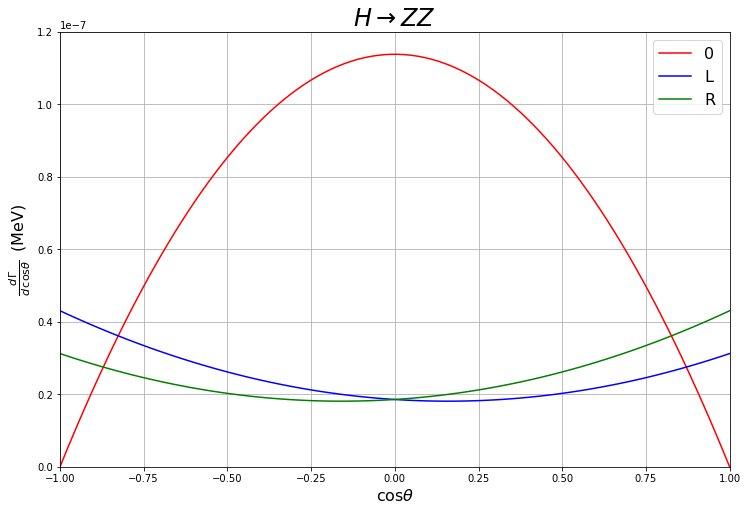}}
\caption{Angular distribution of the positively charged lepton in the $Z$ CoM.}\label{fig:thetal+Z}
\end{figure}

\section{The $H\rightarrow ZZ\rightarrow 4\,\ell$ channel}
\label{sec:ResultsZZ}

Using $M_H = 125.25$ GeV, $M_Z = 91.19$ GeV, $\Gamma_Z =2.50$ GeV, $\sin(\theta_W)^2 = 0.23$,
$\alpha = \frac{1}{127}$ 
 the differential decay width with respect to $\phi$ in $H \rightarrow ZZ$ is
\begin{equation}
\begin{split}
\frac{d \Gamma}{d \phi} =  & \left( 4.216\, f_0^2 + 1.376  \,(f_L^2 + f_R^2) - 
 7.8 \times 10^{-2}\,f_0 \,( f_L  + f_R )\cos\phi  \right. \\
  & \qquad \left. \qquad  + 0.688\,f_L\, f_R \,\cos(2\,\phi) \right) \times 10^{-7}\, 
  \frac{\mathrm{MeV}}{\mathrm{degree}}.
\end{split}
\end{equation}

Taking $f_0 = f_L = f_R =1$ it becomes
\begin{equation}
\label{dGammaDphi_Z}
\frac{d \Gamma}{d \phi} = (6.968 - 0.156\,\cos\phi + 0.688\,\cos(2\,\phi)) \times 10^{-7}\,  
\frac{\mathrm{MeV}}{\mathrm{degree}},
\end{equation}
in good agreement with \rfs{Bredenstein:2006rh,Altenkamp:2018bcs}.
The coefficient of the $\cos(2\,\phi)$ term is about 10\% of the constant one and about four times
larger than the coefficient of the $\cos\phi$ term. The presence of a large contribution proportional to
 $\cos(2\,\phi)$ was pointed out in \cite{Bredenstein:2006rh},
while the smaller term proportional to  $\cos\phi$ went unnoticed.
The longitudinal longitudinal component accounts for about 60\% of the 
partial width while each of the left left and right right components contribute 20\%.

\begin{figure}[!tbh]
\centering
\subfigure{\includegraphics[scale=0.27]{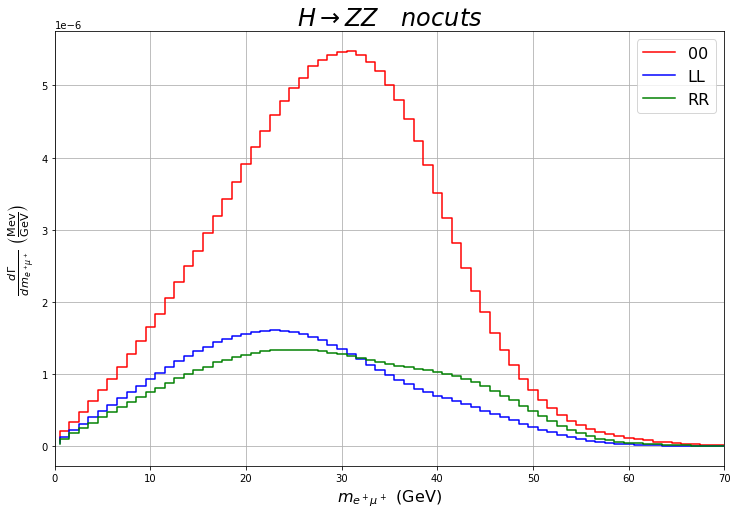}}
\subfigure{\includegraphics[scale=0.27]{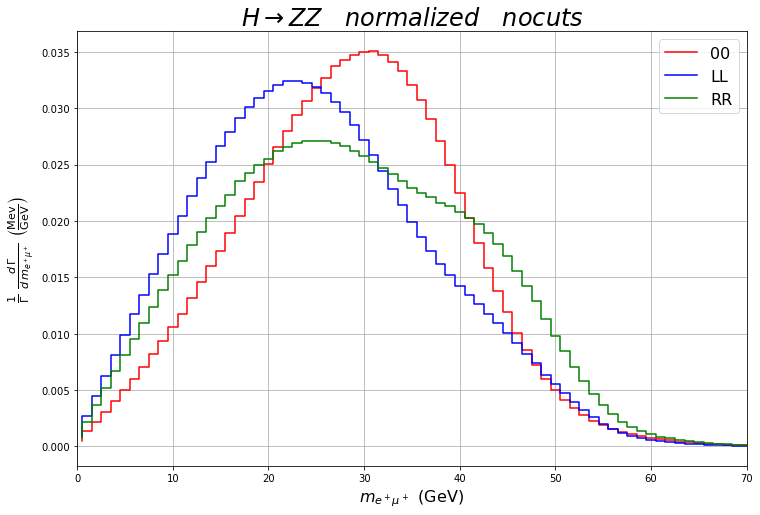}}
\caption{Invariant mass distribution of the $e^+\mu^+$ pairs. No lepton cut is applied. 
The curves on the right are normalized to unit
 integral.}\label{fig:ep_mup_mass}
\end{figure}

\begin{figure}[!tbh]
\centering
\subfigure{\includegraphics[scale=0.3]{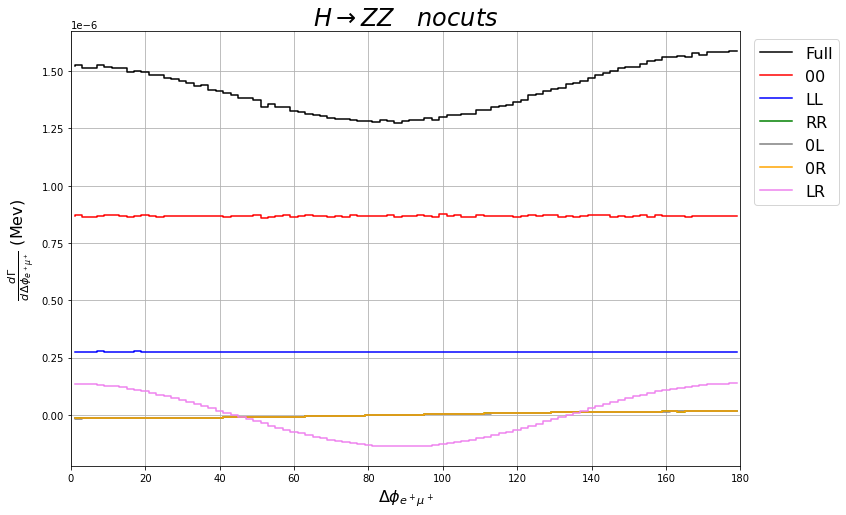}}
\caption{Azimuthal separation, in the Higgs center of mass system, between $e^+$ and $\mu^+$ for $ZZ$ 
events in the absence of cuts. }
\label{fig:DeltaPhi_Z_nocuts}
\end{figure}

One could wonder whether the distributions discussed in this note have any chance of being measured.
We recall that CMS, with 35 $fb^{-1}$ at 13 TeV,  collected about 50 four lepton events on the Higgs peak
\cite{Sirunyan:2019twz,Sopczak:2020vrs}.
For comparison, Run 2 has provided about 140 $fb^{-1}$ to each large experiment; Run 3 is expected to 
accumulate about 200 
$fb^{-1}$ at 14 TeV and finally HL-LHC will deliver 3000 $fb^{-1}$
\cite{CMS:2018mbt,Azzi:2019yne}. Therefore we can expect
of the order of 500 events by the end of Run 3 and thousands of additional events from HL-LHC.

\Fig{fig:Zmass} shows the distribution of the invariant mass of the same flavour, opposite charged leptons.
The curves in the right hand side plot are normalized.
In addition to the expected peak at the $Z$ mass, the curves display a wide increase at small invariant masses.
The secondary peak is wider and extends to smaller value for the longitudinally polarized virtual $Z$'s than 
for the transversely polarized ones. 

\Fig{fig:thetal+Z} presents the distribution of the angle between the positively charged lepton and the 
direction of flight of the $Z$ boson, see \eqn{eq:interfpol}. The longitudinally
 polarized part is distributed as 
$\sin^2\theta$. The right and left terms depend only mildly on the angle, showing a weak preference for the 
forward(backward) direction in the right(left) polarized case. These distributions coincide with the 
familiar ones for on shell $Z$ decay even though in $H \rightarrow ZZ$ each $Z$ is on mass shell only
about 50\% of the times.

In \fig{fig:ep_mup_mass} we study the invariant mass of the two positively charged leptons. 
This quantity has the interesting property of depending
on all five independent variables which describe the decay of the Higgs boson to four fermions.
On the right hand side we show the same curves normalized to unit integral.

The $e^+\mu^+$ invariant mass shows some dependence on the underlying vector boson polarizations:
the longitudinal longitudinal result is more peaked  that the LL and RR ones. It is harder than the LL
distribution.
The RR curve is the widest one, with a tail at larger invariant masses.
The contribution of the interference terms in
\eqn{eq:interfpol} is not zero. However, it is about two orders of magnitude smaller than those in 
\fig{fig:ep_mup_mass} and, therefore, not plotted.
 
\Fig{fig:DeltaPhi_Z_nocuts} shows the azimuthal separation of the two positively charged leptons in the Higgs
center of mass system, with the decay axis in the $z$ direction.  The result agrees with 
\eqn{dGammaDphi_Z}.  The RL interference term provides the bulk of the azimuthal dependence.
The term proportional to $\cos\phi$ is due to the inteference between the longitudinal component and the
R and L ones.
NLO Electroweak corrections for the $\Delta \phi$  differential distribution have been computed in
\rfs{Boselli:2015aha,Altenkamp:2018bcs}. They are about -1\% for $\Delta \phi = \pi$ and +4\%  for $\Delta \phi = 0$.

\section{The \texorpdfstring{$H\rightarrow WW\rightarrow e \mu \nu \nu$ channel}{Lg}}
\label{sec:ResultsWW}

Using  $M_W = 80.38$ GeV, $\Gamma_W =2.10$ GeV, 
 the differential decay width with respect to $\phi$ in $H \rightarrow WW$  is
\begin{equation}
\begin{split}
\frac{d \Gamma}{d \phi} = & \left( 1.762 \,f_0^2 + 0.576\,( f_L^2 +  f_R^2)  + 1.275\, f_0\, (f_L +  f_R) \cos
\phi  \right. \\
  & \qquad \left. \qquad  
 + 0.651\, f_L f_R \cos(2\,\phi) \right) \times 10^{-5}\,  \frac{\mathrm{MeV}}{\mathrm{degree}}.
\end{split}
\end{equation}
Taking $f_0 = f_L = f_R =1$ it becomes
\begin{equation}
\label{dGammaDphi_W}
\frac{d \Gamma}{d \phi} = \left( 2.913 + 2.550 \,\cos\phi + 0.651\,\cos(2\,\phi) \right) \times 10^{-5}\,  
\frac{\mathrm{MeV}}{\mathrm{degree}}.
\end{equation}
\Eqn{dGammaDphi_W}  shows that, in the $W$ case, the coefficient of the $\cos\phi$ term is comparable in 
magnitude with the constant term and about four times larger than the coefficient of the $\cos(2\,\phi)$ one, 
in general agreement with \rf{Altenkamp:2018bcs} which, however, shows a different though related variable, 
the difference in azimuth in the laboratory transverse plane, which is easier to measure. 
Notice that for the negatively charged W,  the $\ell^-$ has opposite three momentum 
in the $W$ rest frame compared to the antiparticle, whose distribution is described in 
\eqn{eq:interfpol}. This implies that for the negatively charged lepton $\phi \rightarrow \pi + \phi$ and 
$\cos\theta \rightarrow -\cos\theta$.
The longitudinal longitudinal component accounts for about 60\% of the 
partial width while the two left left and right right components contribute 20\%.

\begin{figure}[!tbh]
\centering
\subfigure{\includegraphics[scale=0.3]{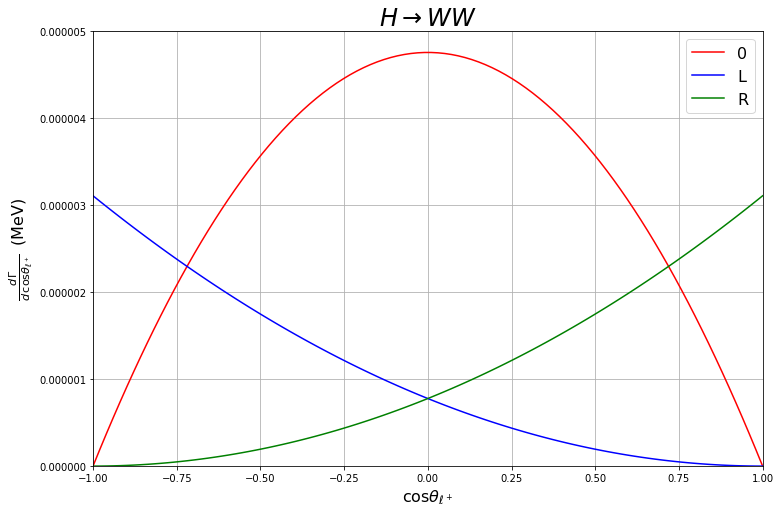}}
\caption{Angular distribution of the positively charged lepton in the $W$ CoM.}\label{fig:thetal+W}
\end{figure}

\Fig{fig:thetal+W} presents the distribution of the angle between the positively charged lepton and the 
direction of flight of the $W$ boson in the reference frame of the latter. 
The longitudinal polarized part is distributed as 
$\sin^2\theta$, while the right and left terms are proportional to $(1\pm\cos\theta)^2$ respectively,
as in on shell $W$ decays.

\begin{figure}[!tbh]
\centering
\subfigure{\includegraphics[scale=0.27]{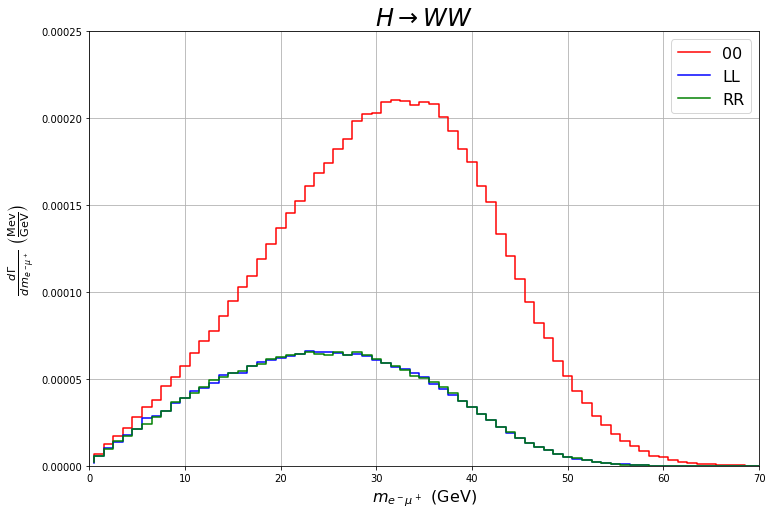}}
\subfigure{\includegraphics[scale=0.27]{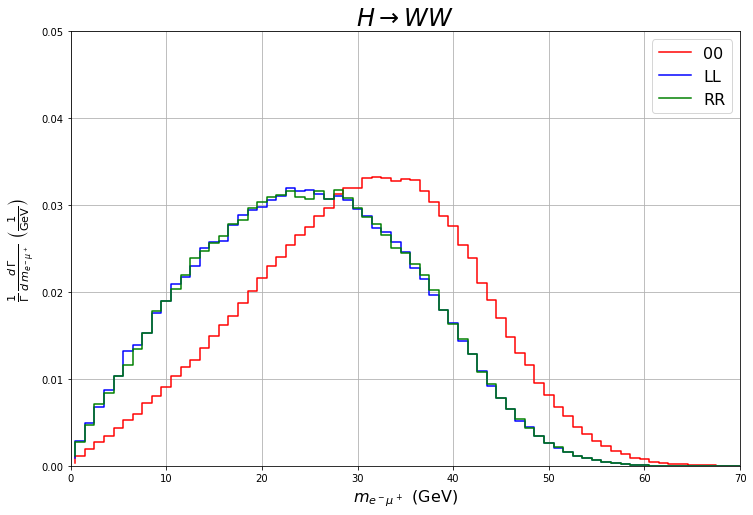}}
\caption{Invariant mass distribution of the $e^-\mu^+$ pair. No lepton cut is applied. 
The curves on the right are
normalized to unit integral.}\label{fig:em_mup_mass}
\end{figure}

\begin{figure}[!tbh]
\centering
\subfigure{\includegraphics[scale=0.3]{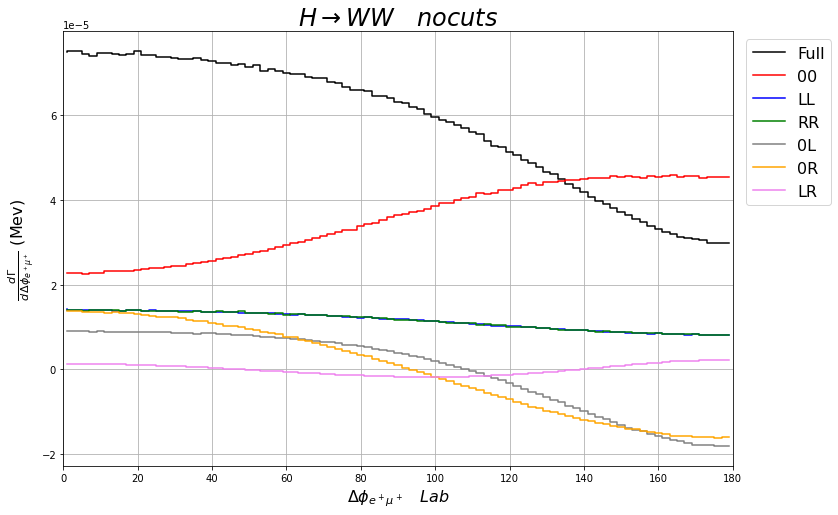}}
\caption{Azimuthal separation, in the laboratory frame,
 between $e^+$ and $\mu^-$ for $WW$ events in the absence of cuts. }
\label{fig:DeltaPhi_Lab_W_nocuts}
\end{figure}

In \fig{fig:em_mup_mass} we study the invariant mass of the two charged leptons.
On the right hand side we show the same curves normalized to unit integral.
In the $W$ case this variable has the additional advantage of 
not requiring the identification of the rest frame of the $W$ pair which is notoriously extremely difficult to
determine because of the two neutrinos in the final state. 
 
The $\ell^-\ell^+$ invariant mass again shows some dependence on the underlying vector boson 
polarizations: the RR and LL curves are identical, as expected, and softer than 
the longitudinal longitudinal result. The contribution of the interference terms in
\eqn{eq:interfpol} are not zero, however, they are about two orders of magnitude smaller than those in 
\fig{fig:em_mup_mass}.

\Fig{fig:DeltaPhi_Lab_W_nocuts} shows the azimuthal separation of the two charged leptons in the laboratory
frame. \Eqn{dGammaDphi_W} shows that in the Higgs center of mass system all diagonal terms are 
independent of the angular separation in the plane orthogonal to the decay axis. In the lab, however,
all polarization combinations depend non trivially 
on the difference in azimuth between the leptons. The full distribution 
favors small separations. At large values of $\Delta\phi$, there is a partial cancellation between the 
longitudinal longitudinal and transverse transverse components, and the longitudinal transverse interferences.
Large interferences in  $\Delta\phi$ have also been reported in  polarized $W^+W^-$ production at the LHC
\cite{Denner:2020bcz}.
NLO Electroweak corrections for the $\Delta \phi$  differential distribution have been computed in
\rf{Altenkamp:2018bcs}. They are about 2.5\% for $\Delta \phi = \pi$ and 3.5\%  for $\Delta \phi = 0$.

\section{A preliminary assessment of the effect of  cuts in the LHC environment}
\label{sec:ResultsCuts}

In this section I investigate whether the differences of the kinematical distributions in the decay of polarized 
vector bosons survive in the LHC environment, where acceptance cuts and additional requirements,
to improve the separation of signal from background, are necessary.
Starting from a sample of $gg \rightarrow H$ events at leading order, the Higgs boson has been 
subsequently decayed to four leptons according to \eqn{eq:interfpol}, 
with a uniform angular distribution of the decay axis. In this simplified setup, all affects due to the transverse 
momentum of the Higgs boson are neglected.

The set of cuts for the $ZZ$ case have been extracted from \rf{Sirunyan:2017exp} by CMS.
\begin{itemize}
\item       $pT_\ell\,>$ 7 GeV, \quad $\vert\eta_\ell\vert\,<$ 2.5   \quad     (acceptance)
\item     12 GeV $ <\,m_{\ell^+\ell^-}\,<$  120 GeV,  \quad  $m_{4\ell}\,>$  70 GeV
\item       $\Delta R_{\ell,\ell}\,>$   0.02,  \quad  $ m_{\ell^+{\ell^\prime}^-}\,>$  4 GeV  \quad 
(veto on soft, collinear pairs)
\item       $N_\ell(pT > 20\, {\mathrm GeV})\,>$  0,  \quad  $N_\ell(pT > 10\, {\mathrm GeV})\,>$   1 \quad 
(high $pT$ leptons)
\end{itemize}

\begin{figure}[!tb]
\centering
\subfigure{\includegraphics[scale=0.3]{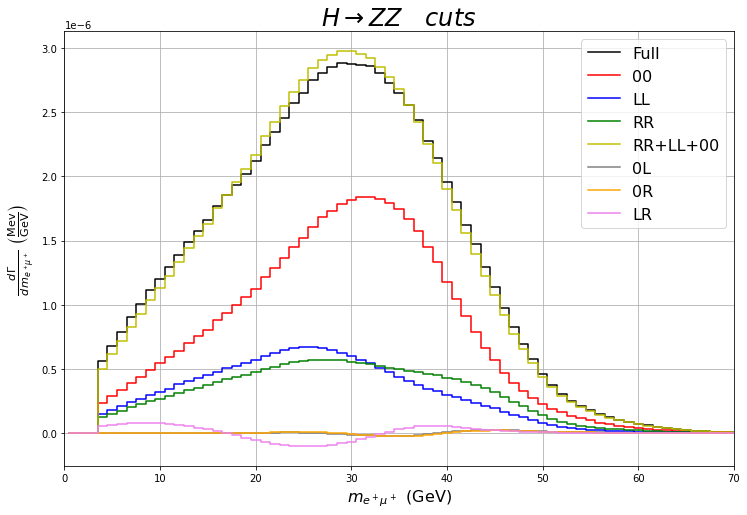}}
\caption{Invariant mass distribution of the $e^+\mu^+$ pairs for 
$ZZ$ events in the presence of cuts.}\label{fig:ep_mup_mass_cuts_Z}
\end{figure}

\Fig{fig:ep_mup_mass_cuts_Z} shows the mass distribution of the $e^+\mu^+$ pairs for each of the six 
combinations of $Z$ polarizations, for the sum of the RR, LL and longitudinal longitudinal contributions and 
for the sum of all contributions.
The shape of the RR, LL and longitudinal longitudinal distributions are very similar to the ones in the inclusive 
case while the normalization decreases by about a factor of three. The LR interference term is small but not 
negligible. Its contribution is positive at the small and large end of the mass range, while it is negative in the 
peak region $m_{e^+\mu^+} \approx 30$ GeV.

\begin{figure}[!tbh]
\centering
\subfigure{\includegraphics[scale=0.27]{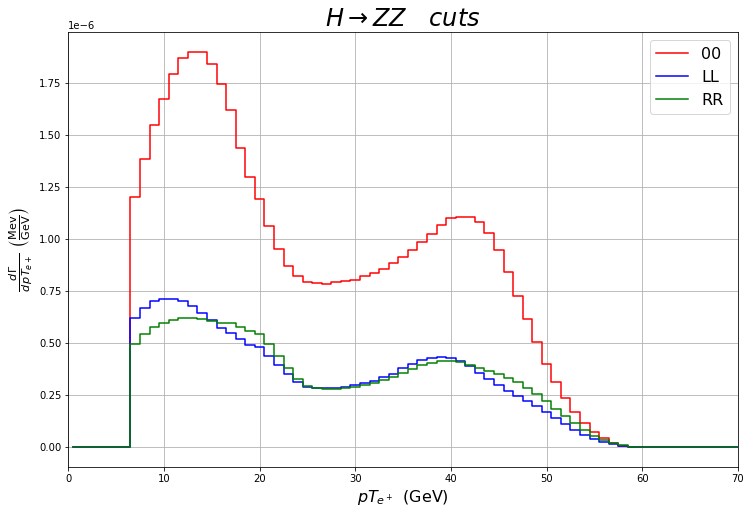}}
\subfigure{\includegraphics[scale=0.27]{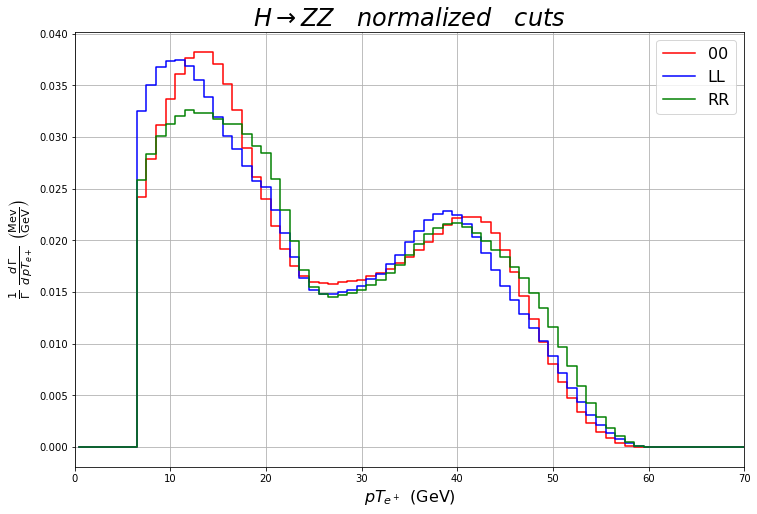}}
\caption{Transverse momentum distribution of the $e^+$ for $ZZ$ events. The curves on the right are
normalized to unit integral. The interference terms are negligible.}\label{fig:ep_pT_Z_cuts}
\end{figure}

The transverse momentum distribution of the $e^+$ for $ZZ$ events is shown in \fig{fig:ep_pT_Z_cuts}
for the RR, LL and longitudinal longitudinal cases. The interference contributions are negligible. The three
distribution show small differences. As expected, they exhibit two broad peaks at about half the value of the   
preferred lepton pair masses in \fig{fig:Zmass}. 

\begin{figure}[!tbh]
\centering
\subfigure{\includegraphics[scale=0.3]{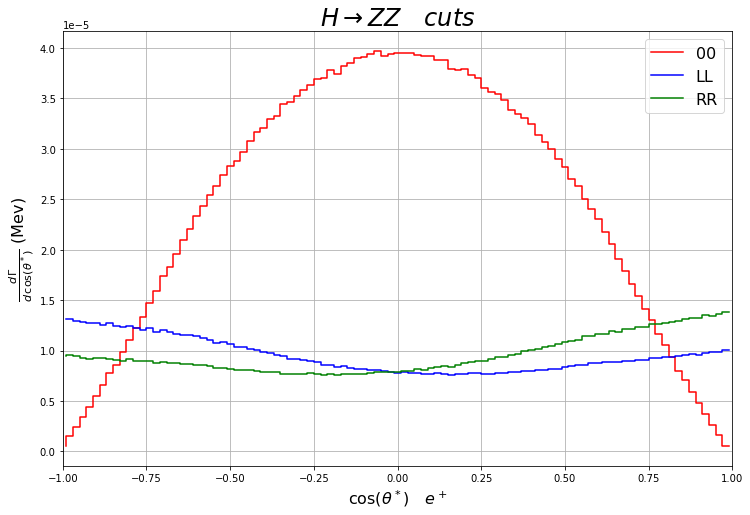}}
\caption{Angular distribution of the positively charged lepton in the $Z$ CoM for the RR, LL and longitudinal longitudinal contributions. The interference terms are negligible.}
\label{fig:thetal+Z_cuts}
\end{figure}

\Fig{fig:thetal+Z_cuts} presents the distribution of the angle between the positively charged lepton and the 
direction of flight of the $Z$ boson for the RR, LL and longitudinal longitudinal contributions. The interference 
terms are negligible. The shape of the distributions are very similar to those in \fig{fig:thetal+Z}. 

\begin{figure}[!tbh]
\centering
\subfigure{\includegraphics[scale=0.3]{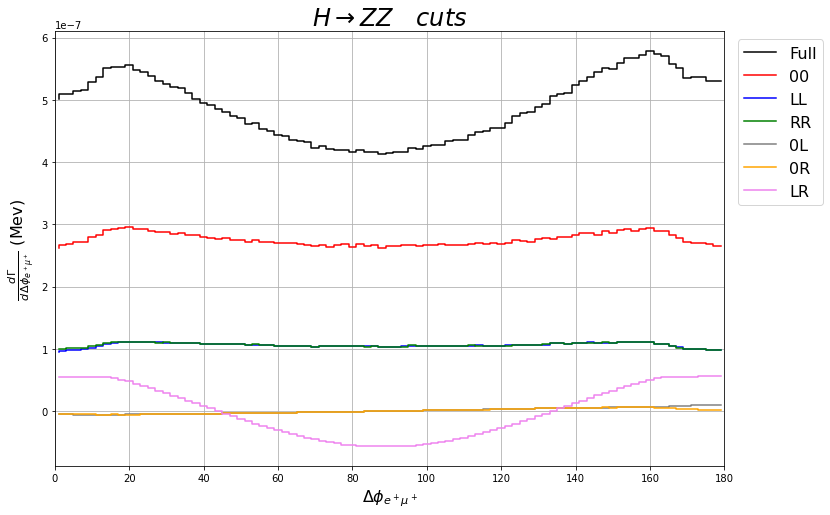}}
\caption{Azimuthal separation, in the Higgs center of mass system, between $e^+$ and $\mu^+$ for $ZZ$ 
events in the presence of cuts. }
\label{fig:DeltaPhi_Z_cuts}
\end{figure}

\Fig{fig:DeltaPhi_Z_cuts} presents the azimuthal separation  between $e^+$ and $\mu^+$ in the plane 
transverse to the Higgs decay axis. The full result is shown alongside the contribution of each polarization 
combination. Most of the curves are almost flat with a modest decrease at $\Delta\phi = \,0,\,\pi$.
The exception is the LR term which behaves basically as $\cos(2\phi)$. The ratio of the different contributions
are are very similar to those in \fig{fig:DeltaPhi_Z_nocuts}.

The set of cuts for the $WW$ case have been taken from \rf{Aaboud:2018jqu} by ATLAS.
\begin{itemize}
\item       $pT_\ell\,>$ 15 GeV, \quad $\vert\eta_\ell\vert\,<$ 2.5   \quad     (acceptance)
\item       10 GeV $<\, m_{\ell^+\ell^-}\,<$  55 GeV,   \quad $N_\ell(pT > 22\, {\mathrm GeV})\,>$  0
\item       $pT_{\ell\ell}\,>$ 30 GeV,  \quad $pT_{miss}\,>$ 20 GeV
\item        $\Delta \phi_{\ell\ell}\,<$   1.8, \quad $\Delta \phi_{(\ell\ell){pT_{miss}}}\,>\, \pi/2$ 
\end{itemize}

\begin{figure}[!tbh]
\centering
\subfigure{\includegraphics[scale=0.3]{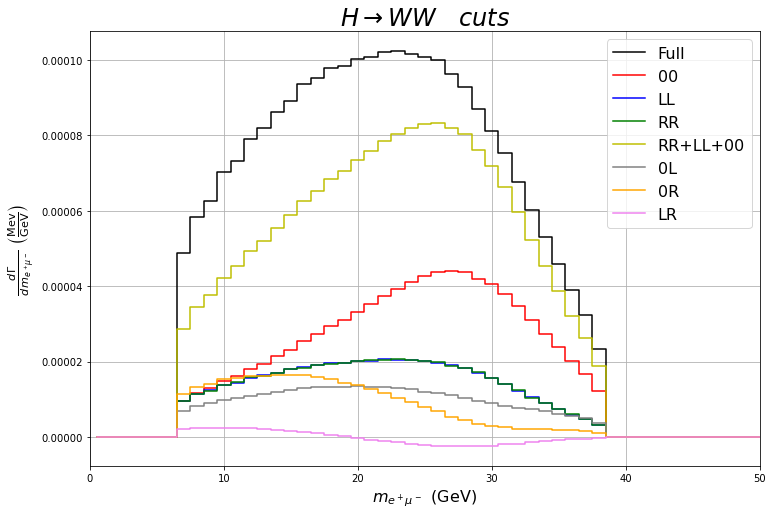}}
\caption{Invariant mass distribution of the $e^+\mu^-$ pairs for $WW$ events in the presence of cuts. }
\label{fig:em_mup_mass_cuts_W}
\end{figure}

\Fig{fig:em_mup_mass_cuts_W} shows the mass distribution of the $e^+\mu^-$ pairs for each of the six 
combinations of $W$ polarizations, for the sum of the RR, LL and longitudinal longitudinal contributions and 
for the sum of all contributions. Contrary to the inclusive case, the interference terms,
particularly those involving one  longitudinal and one transverse $W$, are large, contributing a sizable
fraction of the cross section.

\begin{figure}[!tbh]
\centering
\subfigure{\includegraphics[scale=0.3]{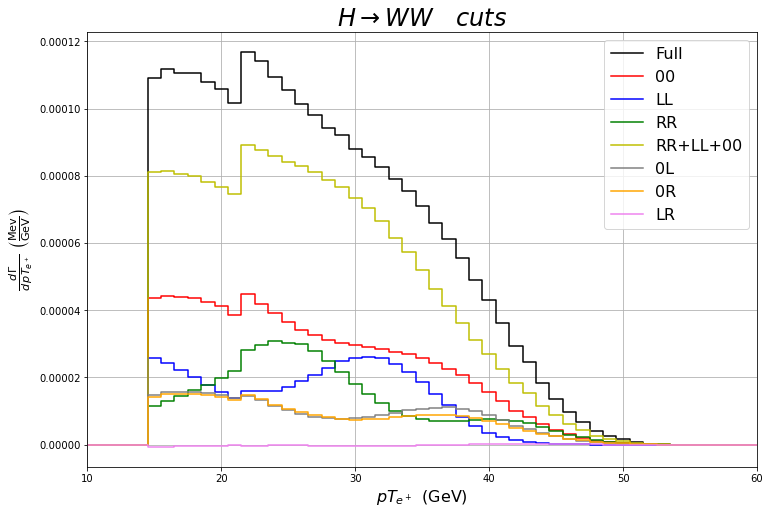}}
\caption{Transverse momentum distribution of the $e^+$  for $WW$ events.}
\label{fig:ep_pT_W_cuts}
\end{figure}

\begin{figure}[!tbh]
\centering
\subfigure{\includegraphics[scale=0.3]{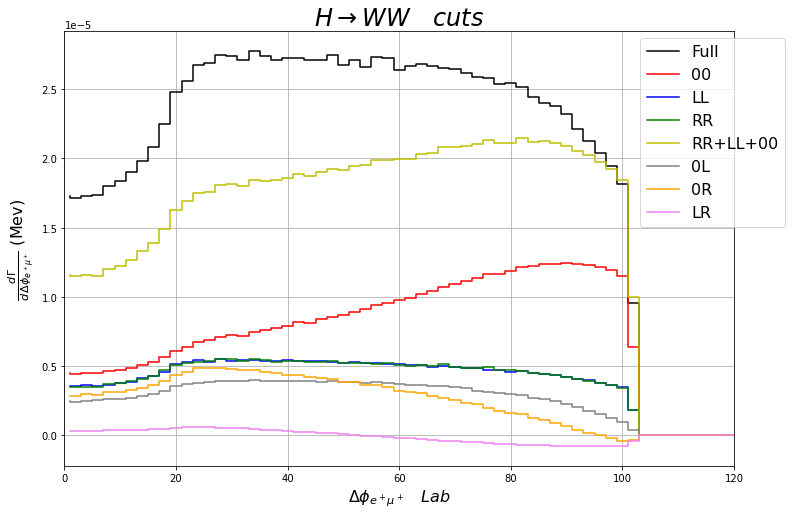}}
\caption{Azimuthal separation, in the laboratory frame,
 between $e^+$ and $\mu^-$ for $WW$ events in the presence of cuts. }
\label{fig:DeltaPhi_Lab_W_cuts}
\end{figure}

The transverse momentum distribution of the $e^+$ for $WW$ events is shown in 
\fig{fig:ep_pT_W_cuts} for each of the six 
combinations of $W$ polarizations, for the sum of the RR, LL and longitudinal longitudinal contributions and 
the sum of all contributions. There is a clear discontinuity at $pT_{e^+} = 22$ GeV related to the requirement 
of at least one charged lepton with such transverse momentum.
The interference terms involving one  longitudinal and one transverse $W$ are large.
The LR contribution is very small.

\Fig{fig:DeltaPhi_Lab_W_cuts} shows the azimuthal separation of the two charged leptons in the laboratory
frame for each of the six 
combinations of $W$ polarizations, for the sum of the RR, LL and longitudinal longitudinal contributions and 
for the sum of all contributions.  The distribution exhibits a sharp drop at  $\Delta \phi_{\ell\ell}\,$=  1.8,
about 103 degrees, due to the veto on back to back leptons. This eliminates the large $\Delta \phi$ region
where the longitudinal transverse interferences are negative.

\section{Conclusions}
In this note I have shown that the amplitude for the Higgs decay to four fermions
can be analytically reformulated in terms of the polarizations of the intermediate vector bosons.
The vector polarizations can be reconstructed
analyzing the kinematic distributions of the final state leptons, providing a new test of the Standard Model.

\section*{Acknowledgements}
Discussions with Alessandro Ballestrero and Giovanni Pelliccioli have been invaluable and are
gratefully ac\-knowledged.\\
The author has been supported by the VBSCan COST Action CA16108 and by the SPIF 
(Precision Studies of Fundamental Interactions) INFN project.

%%%%%%%%%%%%%%%%%%%%%%%%%%%%%%%%%%%%%%%%%%%%%%%%%%%%%%%%%%%%%%%%%%%%%%%%%

% To include bibliography do:
% 1- pdflatex FileName.tex
% 2- bibtex FileName
% 3- pdflatex FileName.tex
% 4- pdflatex FileName.tex
%\bibliographystyle{unsrt}
%\bibliographystyle{hunsrt}
%\bibliographystyle{h-elsevier.bst}

\bibliographystyle{JHEP}

\bibliography{phi}

\end{document}